# Effect of Embedding Watermark on Compression of the Digital Images

Er. Deepak Aggarwal, Er. Kanwalvir Singh Dhindsa

**Abstract -** Image Compression plays a very important role in image processing especially when we are to send the image on the internet. The threat to the information on the internet increases and image is no exception. Generally the image is sent on the internet as the compressed image to optimally use the bandwidth of the network. But as we are on the network, at any intermediate level the image can be changed intentionally or unintentionally. To make sure that the correct image is being delivered at the other end we embed the water mark to the image. The watermarked image is then compressed and sent on the network. When the image is decompressed at the other end we can extract the watermark and make sure that the image is the same that was sent by the other end. Though watermarking the image increases the size of the uncompressed image but that has to done to achieve the high degree of robustness i.e. how an image sustains the attacks on it. The present paper is an attempt to make transmission of the images secure from the intermediate attacks by applying the generally used compression transforms.

——————————— ◆ ———————————

## 1 INTRODUCTION

With the growth of the internet and the immediate availability of computing resources to everyone, "digitized property" can be reproduced and instantaneously distributed without quality loss at basically any cost. The threat to the digitized property has also grown. If we consider the digital image as the digitized property and send it on the network. We have to make sure that it does take much of the bandwidth of the network. That is why we compress the image. Compression may be considered as the attack on the image, means that there could be change image in the process due to intruders on the network. We have to embed the watermark in the image and check for the size of the image and the compression of the image. We have to check that if by embedding some watermark which does not affect the size of the image, we can increase the robustness of the image pertaining to the compression then we better do that. The basic goal is to make the image compression secure by embedding the watermark(s) into the image.

## 2 OBJECTIVES OF THE PAPER

To increase the compression robustness of the digital image by embedding the one or two watermarks keeping the size of the image within the significant limits, as size and security both are the important issues when sending the image on the network. The size is related to the bandwidth and the security is related to the image transmission at the other end with minimum of noise. The watermarked image is then attacked for compression using various transforms. The original image is then compared with watermarked image on the basis of SNR, PSNR and WPSNR.

## 3 THE WATERMARKING PROBLEM

Image watermarking imperceptibly embeds data into a host image. The general process of image watermarking is depicted in figure 1 the original image (Host Image) is modified using the signature data to create the watermarked image. In this process some error or distortion is introduced. To ensure transparency of the embedded data, the amount of image distortion due to the watermark embedding process has to be small. The watermarked image is then distributed and may circulate from legitimate to illegitimate customers. Thereby, it is subjected to various kinds of image distortion. Image distortion may result from e.g. lossy image compression, re-sampling or from specific attacks on the embedded data.

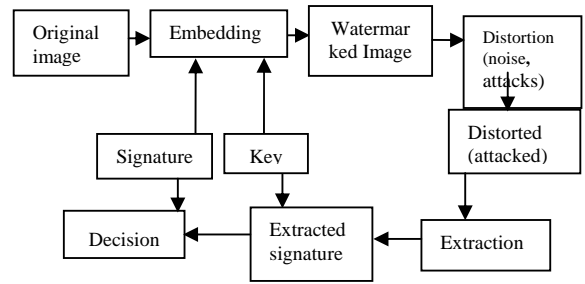

*Figure 1.* A general overview of the data hiding Model

## 4 COMPARISON PARAMETERS

The images that have been regenerated after being compressed or after any other attack can be compared using SNR, PSNR, and WPSNR. Any value of PSNR above 40 will be considered as the good value. This is related to maximum gray level value of any pixel so higher the better. (Ton Kelkar, 2002).

Typical PSNR values range between 20 and 40. They are usually reported to two decimal points (e.g., 25.47). The actual value is not meaningful, but the comparison between two values for different reconstructed images gives one measure of quality. (M. Rabbani and P.W. Jones, 1991)[5]

## 4 CONCEPT OF THRESHOLD

In DIP if have an image having pixel intensity of about 8-10 levels out of which majority of the pixels are having two levels of intensity and we want to reduce the no. of levels to two. So we have select a threshold intensity level between these two levels such that all the pixels having intensity values lesser than the threshold intensity should be assigned the level which equal to the one of the intensity levels that majority pixels are having and all other pixels are assigned the other majority level. Similarly there can be multilevel thresholding in case of images having more objects and more planes. Various pixels can be classified into various object classes depending upon the corresponding threshold value. Thresholding can lead to redundancy and hence more compression. Thresholding can also result in image enhancement in some special case, but that is purely subjective.

## 5 RESULTS AND DISCUSSIONS

TABLE 1: RESULTS OF IMAGE FRUIT. JPG (256X256)

| Image: Fruit.jpg (256×256) | | | | |
|---|---|---|---|---|
| Threshold Level | Transform | SNR | PSNR | WPSNR |
| 50 | DWT2 | 38.7396 | 43.32 | 37.2403 |
| | DCT2 | 50.0031 | 43.33 | 37.1960 |
| | FFT2 | 40.1098 | 43.30 | 36.4787 |
| 100 | DWT2 | 39.0323 | 43.32 | 37.2403 |
| | DCT2 | 54.1222 | 43.33 | 37.2843 |
| | FFT2 | 40.1823 | 43.31 | 36.5933 |
| 150 | DWT2 | 39.3513 | 43.32 | 37.2403 |
| | DCT2 | 56.8139 | 43.33 | 37.3106 |
| | FFT2 | 40.3559 | 43.30 | 36.4750 |
| 200 | DWT2 | 39.7114 | 43.32 | 37.2403 |
| | DCT2 | 58.8964 | 43.34 | 37.3228 |
| | FFT2 | 40.4097 | 43.30 | 36.5017 |
| 250 | DWT2 | 40.4986 | 43.33 | 37.2402 |
| | DCT2 | 62.6829 | 43.36 | 37.3408 |
| | FFT2 | 40.4897 | 43.30 | 36.5607 |
| 300 | DWT2 | 41.6941 | 43.32 | 37.2360 |
| | DCT2 | 65.7207 | 43.85 | 36.8421 |
| | FFT2 | 40.5382 | 43.29 | 36.5338 |

The SNR values are higher for the DCT2 transform otherwise within the range of 40. The values of PSNR are considered to very good if more than 40 dB. Here the maximum value achieved is 43.85 at threshold level of 300. The wPSNR is the having a constant difference with the PSNR so accordingly it is good.

TABLE 2: RESULTS OF IMAGE LENA.JPG (512X512)

| Image:lena.jpg (512×512) | | | | |
|---|---|---|---|---|
| Threshold Level | Transform | SNR | PSNR | WPSNR |
| 50 | DWT2 | 35.7551 | 41.02 | 34.8443 |
| | DCT2 | 46.3525 | 41.19 | 35.0484 |
| | FFT2 | 39.4476 | 41.49 | 34.1719 |
| 100 | DWT2 | 36.3061 | 41.02 | 34.8442 |
| | DCT2 | 48.8283 | 41.11 | 35.0772 |
| | FFT2 | 39.4821 | 41.56 | 34.1166 |
| 150 | DWT2 | 37.2263 | 41.02 | 34.8432 |
| | DCT2 | 50.0269 | 41.14 | 35.1172 |
| | FFT2 | 39.5493 | 41.61 | 34.1028 |
| 200 | DWT2 | 38.3959 | 41.02 | 34.8408 |
| | DCT2 | 50.1241 | 41.14 | 35.1201 |
| | FFT2 | 39.5700 | 41.63 | 34.1106 |
| 250 | DWT2 | 39.7524 | 41.02 | 34.8381 |
| | DCT2 | 51.1136 | 41.13 | 35.1150 |
| | FFT2 | 39.5909 | 41.64 | 34.0665 |
| 300 | DWT2 | 42.0737 | 41.02 | 34.8292 |
| | DCT2 | 54.0281 | 41.07 | 35.0467 |
| | FFT2 | 39.0698 | 41.59 | 34.1313 |

The SNR values are higher for the DCT2 transform otherwise within the range of 40. The values of PSNR are considered to very good if more than 40 dB. Here the maximum value achieved is 41.64 at threshold level of 250 for an image of resolution 512X512.This is also interesting to see that PSNR value may reduce if we increase the threshold level. The wPSNR is the having a constant difference with the PSNR so accordingly it is good.

TABLE 3: RESULTS OF IMAGE DMG.TIFF (64X64)

| Image:dmg.tif (64×64) | | | | |
|---|---|---|---|---|
| Threshold Level | Transform | SNR | PSNR | WPSNR |
| 50 | DWT2 | 40.7863 | 38.47 | 31.4212 |
| | DCT2 | 41.5862 | 37.62 | 29.9113 |
| | FFT2 | 41.5215 | 39.09 | 30.8227 |
| 100 | DWT2 | 40.7863 | 37.32 | 29.1243 |
| | DCT2 | 42.5459 | 38.35 | 30.3635 |
| | FFT2 | 41.5322 | 38.89 | 30.6790 |
| 150 | DWT2 | 40.8262 | 38.23 | 30.1562 |
| | DCT2 | 43.3348 | 38.94 | 30.8103 |
|     | FFT2 | 41.5527 | 38.82 | 30.6080 |
| 200 | DWT2 | 40.8262 | 39.24 | 31.5647 |
|     | DCT2 | 44.0750 | 39.61 | 31.5466 |
|     | FFT2 | 41.5716 | 38.73 | 30.5916 |
| 250 | DWT2 | 40.8262 | 39.07 | 31.2173 |
|     | DCT2 | 44.8015 | 40.48 | 32.6541 |
|     | FFT2 | 41.5992 | 38.70 | 30.5482 |
| 300 | DWT2 | 40.9463 | 38.45 | 30.2367 |
|     | DCT2 | 46.4431 | 40.36 | 32.9730 |
|     | FFT2 | 41.6139 | 38.71 | 30.5646 |

The SNR values are higher for the DCT2 transform otherwise within the range of 40. The values of PSNR are considered to very good if more than 40 dB. Here the maximum value achieved is 40.48 at threshold level of 250. Here the values are little lower as this is a different format having different resolution, hereby signifying that the compression depends upon image format. The wPSNR is the having a constant difference with the PSNR so accordingly it is good.

## 6 CONCLUSION

As far as embedding a single watermark to the digital image is concerned there is very little and insignificant effect on the signal to noise ratio of the original Image and the watermarked image. So, keeping in mind the security issues, it better to do it by embedding the watermark. By cascading the two watermarks one after the other, the robustness of the image increases as we try to compress the image the signal to noise ratio changes significantly. As embedding does not increase the size of the image to a greater extent and the level of robustness it provides pertaining to the compression it is always advised to use double watermark the image when security is primary issue rather than the image size.

## 7 FUTURE ASPECTS

In the thesis, I have considered image compression as the only attack on the image. More attacks such as sharpening, blurring, contrast adjustment and gamma correction etc. The effect of all of them can also be studied and analyzed to draw conclusions. Also the varying images can also be taken for the purpose of multiple attacks. The results can be classified on the basis of the type of the image and so as the conclusions. Second major issue which needs attention is the embedding of multiple watermarks. This means that multiple watermarks can be embedded in the cascading mode and their effect on the image can be studied to draw the conclusions. The same treatment can be done by involving many types of watermarks and their effect to draw the conclusion. Also conclusions can be drawn to know that which watermark has what kind of effect on a particular type of image.

## AUTHORS' INFORMATION

First Author: Er. Deepak Aggarwal, presently working as Lecturer in the Department CSE/IT of BBSB Engineering College, Fatehgarh Sahib,Punjab (INDIA). He is having a total teaching experience of about 7 years & presently doing Master of Technology from Punjab Technical University. His major research interests include DIP and performance evaluation of networks. Also Deepak Aggarwal is having to his credit about 7 publications in various National and International Conferences and Journals.

Second Author: Er. Kanwalvir Singh Dhindsa, presently working as Assistant Professor, Department of CSE/IT, BBSB Engineering College, Fatehgarh Sahib, Punjab, India. He is having a total teaching experience of about 10 years and presently pursuing Ph.D from Punjabi University Patiala. His major research interests include Data Bases Modelling Languages and Software Engineering. Also Prof. Dhindsa is having to his credit about 15 publications in various National and International Conferences and journals.